\documentstyle[aps,amssymb,preprint]{revtex}
\begin{document}
\draft
\tightenlines
 
\title{Series expansion for a stochastic sandpile}
\author{J\"urgen F. Stilck$^{1,*}$, Ronald Dickman$^{2,\dagger}$ and
Ronaldo R. Vidigal$^{2,\ddagger}$}
\address{
$^1$Departamento de F\'{\i}sica,
Universidade Federal Fluminense,
Av. Litor\^anea s/n,\\ 
24210-340\\
Niter\'oi - Rio de Janeiro, Brasil\\
$^2$Departamento de F\'{\i}sica, ICEx,
Universidade Federal de Minas Gerais,\\
30123-970
Belo Horizonte - Minas Gerais, Brasil\\
}

\date{\today}

\maketitle
\begin{abstract}
Using operator algebra, we extend the series for the 
activity density in a one-dimensional stochastic 
sandpile with fixed particle density $p$,
the first terms of which were obtained via perturbation theory
[R. Dickman and R. Vidigal,
J. Phys. A {\bf 35}, 7269 (2002)].  The expansion is in
powers of the time; the coefficients are polynomials
in $p$.  We devise an algorithm for evaluating expectations of operator
products and extend the series to ${\cal O}(t^{16})$.
Constructing Pad\'e approximants to a suitably transformed series,
we obtain predictions for the activity that compare well against
simulations, in the supercritical regime. 
\end{abstract}
\vspace{1cm}


\noindent {\small $^*$electronic address: jstilck@if.uff.br}   \\
\noindent {\small $^\dagger$electronic address: dickman@fisica.ufmg.br} \\
\noindent {\small $^\ddagger$electronic address: rvidigal@dedalus.lcc.ufmg.br}

\newpage

\section{Introduction}

Sandpiles with a strictly conserved particle density (so-called
{\it fixed-energy sandpiles} or FES \cite{dvz}), exhibit
absorbing-state phase transitions \cite{marro,hinrichsen,bjp00}, 
and have recently attracted much interest.
Until now, most quantitative results for FES have been 
based on simulations \cite{fes2d,rossi,manna1d,mnrst}, an 
important exception being the solution by Priezzhev {\it et al.}
\cite{priezzhev} of a directed, fixed-energy version of the Maslov-Zhang
model \cite{mz}, via the Bethe ansatz.
It is therefore of great interest to develop theoretical approaches
for FES.  

Series analysis has proved to be one of the most accurate
and reliable approaches to critical phenomena, in both equilibrium  
and nonequilibrium contexts \cite{baker,rdjsp,jsp2,iwandp}.
Series expansion typically functions best in low-dimensional systems
(because longer series can be derived), that is, for just those systems
in which the renormalization group and expansion about an upper
critical dimension $d_c$ are less reliable.  In the case of sandpiles,
systematic epsilon expansions are as yet unavailable, and the value
of $d_c$ in fact remains controversial \cite{pmb,socprl,wijland}.
Simulation results suggest novel critical behavior in the one-dimensional
FES, although conflicting critical exponent values have been
reported \cite{manna1d,mnrst,lubeck,chate}, which may reflect
finite-size effects.  Series expansions, on the other hand, implicity
treat the infinite-size limit, and so provide important information, 
complementary to that afforded by simulations.  In light of these
observations, we believe it highly desirable to apply series methods
to sandpile models.

This paper is one of a series analyzing
a stochastic sandpile using operator methods.  In an earlier work \cite{manft}
a path-integral representation was developed and 
an expansion derived for the order parameter (activity density) in powers of
time.  While the path-integral formalism reveals interesting features of
the model, and may be applied in any number of dimensions, the
complexity of the diagrammatic expansion limits the
number of terms that can be obtained.  (In Ref. \cite{manft} terms
up to ${\cal O}(t^5)$ are reported.)
In this paper we employ a different approach, which
permits us to extend the series for the one-dimensional case 
considerably.  After
casting the master equation for the sandpile in terms of
an operator formalism, we analyze the direct expansion of
its (formal) solution, leading to an 
algorithm for generating the series coefficients.

We consider Manna's stochastic sandpile in its fixed-energy 
(particle-conserving) version \cite{manna1d,manft,manna,manna2}.   
The configuration is specified by the occupation number
$n$ at each site; sites with $n \geq 2$ are said to be 
{\it active}, and have a positive rate of
{\it toppling}.  When a site topples, it loses exactly two
particles (``grains of sand"), which move randomly and 
independently to nearest-neighbor (NN) sites.
(Any configuration devoid of active sites is {\it absorbing},
i.e., no futher evolution of the system is possible once such
a configuration is reached.)
In this work, as in Ref. \cite{manft}, we adopt a toppling rate of
$n(n\!-\!1)$ at a site having $n$ particles, which leads us to define the
order parameter as $\rho = \langle n(n-1) \rangle$. 
While this choice of rate represents a slight departure from the
usual definition (in which all active sites have the same toppling rate), 
it leads to a much simpler
evolution operator, and should yield the same
scaling properties \cite{manft}. 
Preliminary simulation results \cite{rdunp}
indicate that in one dimension the model exhibits a continuous phase transition
at $p_c \!=\! 0.9493$.  

In the following section we define the model and review the operator 
formalism introduced in Ref. \cite{manft}.  This is followed in Sec. III
by an analysis leading to an expansion in terms of so-called reduced commutators.
Implementation of the expansion in a computational algorithm is described 
in Sec. IV.  Then in Sec. V we report numerical results of the series analysis.
A summary and discussion is provided in Sec. VI.

\section{Model}

As discussed in Ref. \cite{manft}, the master equation for this model
may be written in the form
\begin{equation}
\frac{d|\Psi\rangle}{dt} = L |\Psi\rangle \;,
\label{me}
\end{equation}
where
\[
|\Psi \rangle = \sum_{\{n_i\}} p(\{n_i\},t) |\{n_i\} \rangle,
\]
is the probability distribution, and the 
evolution operator
takes the form, 
 \begin{equation} 
L = \sum_i \left[ \frac{1}{4} (\pi_{i-1} + \pi_{i+1})^2 
 - \pi_i^2 \right] a_i^2  \equiv \sum_i L_i.
\label{evop} 
\end{equation} 
Here $a_i$ and $\pi_i$ are, respectively, annihilation and creation
operators associated with site $i$, defined via 
\[
a_i |n_i\rangle = n_i |n_i\!-\!1\rangle
\]
and
\[
\pi_i |n_i\rangle = |n_i\!+\!1\rangle .
\]
The formal solution of the master equation is $|\Psi (t)\rangle = e^{tL}|\Psi (0)\rangle $;
that for the activity density is:
\begin{equation}
\rho(t) = \langle \;|a_0^2 e^{tL} |\Psi (0)\rangle.
\label{act}
\end{equation}
Here we have introduced the notation:

\begin{equation}
\langle \;| \equiv \sum_{\{n\}} \langle \{n\} |
\end{equation}
for the projection onto all possible states;  thus normalization reads:
$\langle \;| \Psi \rangle = 1$.
We consider a uniform Poisson-product initial distribution.
Letting $p_n = e^{-p} p^n/n!$, and using $|P \rangle_i =
\sum_{n_i} p_{n_i} |n \rangle_i$ to denote a Poisson distribution
at site $i$, we have,
\begin{equation}
|\Psi (0)\rangle = \prod_j  |P \rangle _j \;.
\label{psi0}
\end{equation}
We shall expand equation (\ref{act}) for the activity density
in powers of $t$. 

\section{Operator Algebra}

To begin we note some basic properties of operators
$a_j$, $\pi_j$ and $L_j$:

\begin{equation}
a_j^n |P\rangle_j = p^n |P\rangle_j
\label{aP}
\end{equation}

\begin{equation}
\langle \;| \pi_j = \langle \;|
\label{normpi}
\end{equation}

\begin{equation}
\langle \;| L_i = 0
\label{Lcons}
\end{equation}

\begin{equation}
[a_i,\pi_j] = \delta_{i,j}
\label{comm}
\end{equation}

\begin{equation}
[L_i,L_j] = [a_i,L_j] = 0 \;\;\; \mbox{for} \; |i-j| > 1       \;.
\label{comm0}
\end{equation}
The second relation expresses the fact that the creation operator conserves
the normalization of any state, while the third shows that 
$L_i$ conserves probability, as it must.

The coefficient of $t^n/n!$ in the expansion of the activity is:

\begin{equation}
\sum_{\cal S} \langle \;|a_0^2 L_{s_1} L_{s_2}...L_{s_n} |P\rangle \;,
\label{coefn}
\end{equation}
where the sum is over all sequences ${\cal S}$ of sites $s_0 \equiv 0,s_1,,,s_n$ with
$|s_1| \leq 1$, and $s_{j+1} \in \{s_{j,min}-1,...,s_{j,max}+1\}$,
for $j \geq 1$, where $s_{j,min} = \min \{s_0,...s_j\}$,
and $s_{j,max}$ is the maximum of this set.
The restriction on sequences follows from equations (\ref{Lcons}) and (\ref{comm0});
if the condition were violated, it would be possible to move one of the
$L_j$ to the left of all other operators, yielding a result of zero.

Our strategy for evaluating $\rho_{\cal S}$
is to commute each $L_j$ to the left of $a_0^2$.  The first step
replaces $a_0^2 L_{s_1}$ by its commutator, due to
equation (\ref{Lcons}).  If we write this commutator in 
{\it normal order}, that is, with all creation operators $\pi_j$ to the
left of all annihilation operators, then the
$\pi$'s may be replaced by 1, by equation (\ref{normpi}).
Thus,
\begin{equation}
\langle \;| a_0^2 L_j = \langle \;| [a_0^2,L_j]_R \;,
\label{troca}
\end{equation}
where the subscript $R$ denotes a {\it reduced commutator},
that is, the commutator in normal order, with all $\pi$'s replaced by 
unity.  Evidently $[a_0^2,L_j]_R$ involves only annihilation operators.
The two nontrivial expressions of this kind are:

\begin{equation}
[a_0^2,L_0]_R = -2a_0^2 - 4a_0^3 \;,
\label{cr00}
\end{equation}
and 
\begin{equation}
[a_0^2,L_1]_R = \frac{1}{2}a_1^2 + 2 a_0 a_1^2 \;.
\label{cr01}
\end{equation}

In the computational algorithm, discussed in some detail below, it is not
necessary to generate the tree structure of sequences explicitly, since each
monomial is processed separately and both translation and reflection
symmetries may be used in the calculations of the contributions $\rho_{\cal
S}$. Evaluating the expectation of each term in $\rho_{\cal S}$ is trivial,
because

\begin{equation}
\langle \;| a_{s_1}^{m_1}...a_{s_n}^{m_n}|P\rangle = p^M, 
\label{exppa}
\end{equation}
where 
$M = \sum_j m_j$ is the number of annihilation operators, irrespective
of which sites are involved.

It remains to find a general expression for the reduced commutator
$[F^{(j)},L_k]_R$.  Since $F^{(j)}$ is a linear combination of 
products of annihilation operators, and recalling that $a_i$ and $L_k$
commute if $|i \!-\! k| > 1$, we see that the problem reduces to
evaluating

\begin{equation}
C(p,q,r) \equiv [a_{-1}^p a_0^q a_1^r, L_0]_R \;.
\end{equation}
(Commutators involving $L_j$ with $j \!\neq \! 0$ are obtained using
translation invariance.)
It is straightforward to evaluate $C(p,q,r)$ using the following identities.
First we note that

\begin{equation}
[a_j^p, \pi_j] = pa_j^{p-1} \;,
\label{id1}
\end{equation}
as is readily shown by induction.  Using this it is simple to
show 
\begin{equation}
[a_j^p, \pi_j^2]_R = p(p\!-\!1) a_j^{p-2} +2pa_j^{p-1} \;,
\label{id2}
\end{equation}
and
\begin{equation}
[a_j^p, \pi_j^2 a_j^2]_R = p(p\!-\!1) a_j^p +2pa_j^{p+1} \;.
\label{id3}
\end{equation}
Finally, we may use equation (\ref{id1}) to show that for $i \! \neq \! j$,

\begin{equation}
[a_i^p a_j^r, \pi_i \pi_j]_R = pa_i^{p-1} a_j^r + r a_i^p a_j^{r-1}
  + pr a_i^{p-1} a_j^{r-1} \;.
\label{id4}
\end{equation}
Applying these relations one readily finds:

\begin{eqnarray}
\nonumber
C(p,q,r) &=& a_0^{q+2} \left[ \frac{1}{4}p(p\!-\!1)a_{-1}^{p-2} a_1^r
+ \frac{1}{4}r(r\!-\!1)a_{-1}^p a_1^{r-2} \right.
\\
\nonumber
&+& \left. \frac{1}{2}pr a_{-1}^{p-1} a_1^{r-1}
+ p a_{-1}^{p-1} a_1^r  + r a_{-1}^p a_1^{r-1} \right]
\\
&-& q a_{-1}^p  a_1^r \left[ 2 a_0^{q+1} + (q\!-\!1) a_0^q \right] \;.
\label{cpqr}
\end{eqnarray}
Using this result, we can evaluate the reduced commutators 
in a computer algorithm.

\section{Computational Algorithm}

Let us discuss some details of the computer algorithm used to generate the
series for the activity. We employ a recursive procedure to generate the
contributions of order $n+1$ on the basis of those of order $n$. From equation
(\ref{act}) and with a Poisson-product initial distribution 
defined in equation
(\ref{psi0}) we notice that
\begin{equation}
\frac{d \rho}{d t}=\langle \, |a_0^2 \, L \, e^{tL}|P\rangle=
\langle \, | [a_0^2,L]_R \, e^{tL}|P\rangle=-\rho+\langle \,
|(4a_0a_1^2-4a_0^3) \, e^{tL}|P\rangle.
\label{drho}
\end{equation}
The last equality above may be understood using the reduced commutators
(\ref{cr00}) and (\ref{cr01}). Using reflection symmetry, we have
$\langle \, | [a_0^2,L]_R \, e^{tL}|P\rangle = \langle \, |
\{[a_0^2,L_0]_R+2[a_0^2,L_1]_R \} e^{tL}|P\rangle$ and further simplification
is provided by translation symmetry. The coefficient of $t^n/n!$ in the
expansion of the activity may be identified with $C_n$, where 
\begin{equation}
C_n=\left. \frac{d^n \rho}{d t^n} \right|_{t=0}. 
\label{cdef}
\end{equation}
Using the procedure described above we obtain a recursion relation
\begin{equation}
 C_{n+1}=-C_n+\langle |F_{n+1}|P\rangle,
\label{rr}
\end{equation}
where $F_{n+1}=[F_n,L]$ and $F_1=4a_0a_1^2-4a_0^3$. To complete the algorithm,
we have that $C_0(0)=\rho(0)=\langle \, |a_0^2| \rangle=p^2$, where relation
(\ref{exppa}) is used. An immediate
consequence of these recursion relations is that the coefficient of $p^2$ in
the term of order $n$ in $t$ is given by $(-1)^n$, since each monomial in the
functions $F$ has at least three annihilation operators\cite{note1}. This was already
shown in Ref. \cite{manft}.

The calculations were done in two steps. Initially, the functions $F$ were
calculated up to order 12. Each monomial was represented by three integer
variables: an eight-byte integer for the numerator of the coefficient, 
a four-byte integer for the denominator (which is always a power of 2), 
and another eight-byte
integer to store the number of factors of each annihilation operator. 
Since all calculations are done in integer arithmetic, there are no roundoff
errors. Using 
translation invariance, each monomial was put in a form such that the 
the annihilation operator of lowest spatial index is $a_0$. 
The power $m_i$ associated with each
annihilation operator $a_{s_i}$ (as in equation (\ref{exppa}))
is stored in four bits of the eight-byte integer variable 
mentioned above. As each new monomial is generated, a search is performed 
for any existing term with the
same set of powers; storing all powers in a single integer facilitates
the search.  As a consequence of equation (\ref{comm0}), when the
reduced commutator of a monomial with $L$ 
is calculated, nonzero contributions may arise only from the commutator 
of the monomial with $L_{-1}$, $L_0$, $L_1, \ldots, L_{i+1}$, where $i$ 
is the largest index in the monomial. As is clear 
from equation (\ref{cpqr}), 
each of these commutators can give rise to up to seven new monomials. 
Thus, it is apparent that the number of monomials grows very rapidly 
as the order is increased; the function $F_{12}$ involves $519 \; 115$ 
monomials. To go beyond order 12 it is
necessary to handle integers larger than can be represented using eight
bytes and to process monomials with more than 16 exponents,
which can no longer be stored in a single 8-bite integer
variable. In fact, at order 12 most of the processing time is used in the
search procedure. Therefore, our results from order 13 to 16 were obtained
processing the monomials in $F_{12}$ one-by-one, generating all
contributions from it at orders 13-16. In these calculations the
numerators were represented by two eight-byte integers. The limiting 
order (16) is determined by the large number of new monomials
generated; a single monomial in $F_{15}$ may generate on the order of 40 
monomials in $F_{16}$. 
The results presented here required about 170 hours of cpu time on an
Athlon K7 1800 MHz computer.

It is convenient to write the expansion in the form:
\begin{equation}
\overline{\rho}(t) \equiv \frac{\rho(t)}{p^2} 
= \sum_n \frac{(-t)^n}{n!} \sum_{m=0}^{n-1} b_{n,m} p^m \;.
\label{bnm}
\end{equation}
The series coefficients $b_{n,m}$ are listed in Table I.  In Ref. \cite{manft} it was shown that
$m \leq n\!-\!1$, with
\begin{equation}
b_{n,n-1} =
2^{4n-1} \frac{(2n-1)!!}{(2n+2)!!}.
\label{last}
\end{equation}
The coefficients reported in Table I satisfy this relation at each order,  
and agree with those derived (for $n \leq 5$) 
using the path-integral formalism \cite{manft}.

\section{Analysis of Series}

The coefficients $b_{n,m}/n!$ in the time series, equation (\ref{bnm}), 
grow rapidly with $n$; the rate of growth appears to be faster than
exponential.  This is evident from an analysis of
\begin{equation}
h_{n,m} \equiv \ln \left( \frac{b_{n,m}}{n!} \right)
\label{deff}
\end{equation}

To see if the $h_{n,m}$ follow a systematic trend we analyze these quantities
for a given $q \equiv (m-1)/(n-2)$.  (For a fixed value of $n$, the 
$h_{n,m}$ appear to trace out a smooth curve, so that $h_n(q)$ for 
intermediate values of $q$ can be estimated via interpolation.) 
As shown in figure 1, $h_n(q)$ appears to grow faster than exponentially
with $n$, away from the limits $q=0$ and $q=1$.  
(Observe that for $m = 0$,
$h_{m,n} \to - \infty$ as $n \to \infty$
since $b_{n,0} = 1$, and similarly for $m=n-1$, since
equation (\ref{last}) implies that $b_{n,n-1}$ grows more slowly than $n!$.)
A reasonable description of the dominant growth in the series 
coefficientes is $h_n(q) \sim n^\alpha(q)$, with the exponent $\alpha$
(see the inset of figure 1) taking its maximum value of about 1.2 
for $q \simeq 0.4$.  This of course implies faster-than-exponential growth
for the coefficients $b_{n,m}/n!$. 

Next we examine the behavior of the coefficients in 
the time series for specific values of the particle density $p$.
Let 
\begin{equation}
\overline{\rho}(t) = \sum_n c_n (-t)^n
\label{rhot}
\end{equation}
where
\begin{equation}
c_n \equiv \frac{1}{n!} \sum_{m=0}^{n-1} b_{n,m} p^m \;.
\label{cn}
\end{equation}
For $p=1$ (slightly above the critical value of 0.9493), $c_n$ is 
simply the sum of all
coefficients at order $n$, divided by $n!$.  
The coefficients $c_n$ again grow faster than exponentially,
with $\ln c_n \sim n^{1.15} $ for $p=1$ and $\sim n^{1.10}$
for $p=2$.  (Given the limited number of coefficients, we cannot
make very precise estimates of the exponent.  The key point is that
the growth appears to be faster than exponential.)
These results imply that equation (\ref{rhot}) is
a {\it divergent series} with zero radius of convergence.

We turn now to an analysis of the series for $\overline{\rho}(t)$.
As is well known, it is often possible to obtain useful results 
from divergent series by means of a resummation technique.
In the present case we construct Pad\'e approximants to the
time series or to the series obtained via a transformation of 
variable \cite{jsp2,baker75,orszag,guttmann}.
We have examined many transformations, for example
\begin{equation}
y = \frac{1 - e^{-bt}}{b} \;,
\label{exp}
\end{equation}
\begin{equation}
x = \frac{t}{1+bt} \;,
\label{euler}
\end{equation}
\begin{equation}
z = 1 - \frac{1}{(1+bt)^\gamma} \;,
\label{gamma}
\end{equation}
\begin{equation}
w = 1 - \frac{1}{1+ \ln(1+bt)} \;,
\label{ln}
\end{equation}
and
\begin{equation}
v = 1 - \exp \left\{b\left[1-(1+t)^\gamma \right] \right\} \;.
\label{ste}
\end{equation}

Each transformation maps the interval $t \geq 0$ to 
a finite interval, and can be expanded as a power series in $t$ about
$t=0$, with the lowest-order term $\propto t$.  Each
is readily inverted permitting one to express the time $t$
in powers of the new variable.  
The slow convergence associated with the power-law or logarithmic
forms in the last four expressions is motivated by the
numerical finding of slow relaxation in the sandpile model,
even far from the critical point \cite{mannrel}.  
We analyze the transformed
series for $\overline{\rho}$ (or for $\ln \overline{\rho}$) using
Pad\'e approximants.
The degree of success
depends greatly on the range of $p$ under consideration.
(Each transformation includes a free parameter $b$, which can be adjusted
to optimize the regularity of the result, or to obtain consistency between
different approximants.  Except where noted, 
the results do not exhibit much sensitivity 
to the choice of this parameter.)

For small values of $p$, the best results are obtained via Pad\'e 
approximants to the $t$-series {\it without} any transformation
of variable.  Figure 2 compares 
series predictions for $p=0.5$ (obtained using the [6,7], [7,8]
and [7,9] approximants
to the series for $\overline{\rho}(t)$)
against the result of a Monte Carlo simulation for a 
system of 500 sites (for $p=0.5$ finite-size effects are negligible at this 
system size).  The [7,8] approximant is reliable for
$t \leq 10$.  (Various other approximants, such as [8,8] and [7,7], are ill-behaved
and provide reasonable predictions only for quite short times, typically $t \leq 2$.)
We have not been able to improve the series prediction for  
longer times, either by a change of variable or through analysis
of $\ln \overline{\rho}(t)$ or its time derivative.  Although some
improvement could be expected with longer series, it appears
unlikely that the asymptotic decay of $\overline{\rho}(t)$ in the
subcritical regime will
be accessible through analysis of an expansion in powers of time.

For larger values of $p$ the transformation defined in equation
(\ref{gamma}), using $\gamma = 1/2$, 
is the most useful of those studied.  In figure 3
we compare the [8,8] approximant (obtained using 
$b = 0.57$) with simulation data for $p=1$.  The situation is markedly better
than for $p=1/2$: the series prediction accompanies
the simulation result up to around $t=1000$.  It must be noted, however, 
that the good agreement seen here depends on the choice of the 
transformation parameter
$b$.  For other values the agreement with simulation is not as good.
(A more suitable criterion for choosing $b$ would be by seeking 
agreement among various approximants \cite{jsp2}.  In the present case 
this is not possible because the off-diagonal approximants to the $z$ 
series are ill-behaved, while the [7,7] approximant behaves very 
similarly to the [8,8] used here.)  Despite the good agreement up to
times on the order of 1000, the present series seems incapable of capturing
the asymptotic long-time relaxation of $\overline{\rho}(t)$, which is
non-monotonic, as shown in figure 3.

Remarkably, the reliability of the series improves dramatically 
at larger values of the particle density $p$.  Series 
and simulation results for $\overline{\rho}(t)$ at $p=2$ are compared
in figure 4; the maximum relative error is about 0.1\%.  
(The series prediction is generated as for $p=1$, but using
$b=1.5$ in this case.)   The good agreement, moreover, persists
at long times, motivating a study of $\overline{\rho}_\infty
\equiv lim_{t \to \infty} \overline{\rho}(t)$, corresponding to
the transformed series with $z=1$ in equation (\ref{gamma}).
(We again use the [8,8] approximant to the $z$-series.)  

The series prediction for $\overline{\rho}_\infty$
is compared with simulation in figure 5, using parameters $b=0.57$ and $b=5$. 
Excellent agreement is found for
$p \geq 2$, the relative error being $\leq 0.2\%$.
The smaller $b$ value yields better results for $p \simeq 1$, whereas
slightly better results are obtained for larger $p$, using the larger
$b$ value.   For $p \geq 2$ we may claim quantitative accuracy
for the series prediction.  Nearer the critical point, the agreement
appears reasonable (at least on the scale of figure 5), but it is
clear that the 16-term series cannot be used to study critical
properties.   For example, the prediction using $b=0.57$ yields a
critical value of about 0.906, that is, the extrapolated activity density
goes to zero at this $p$ value.  The critical value found in simulations
is 0.9493.

In summary, the present series seems quite reliable in the
supercritical regime, both at short and at asymptotically long times,
whereas its utility in the critical and subcritical regimes is restricted
to rather short times.   Just above the critical point, rather good
predictions are possible for short and intermediate times, but this
depends on a judicious choice of the transformation parameter $b$.

\section{Discussion}

We develop an algebraic method leading to a time series
for the activity density of the stochastic sandpile model
introduced in \cite{manft}.  Determination of the series coefficients
depends on evaluation of certain commutators, an algebraic task
readily codified in a computational algorithm.  We extend the
series for the one-dimensional case to sixteen terms.

Analysis of the series yields disappointing results for the
subcritical and critical regimes, but very good predictions in the
supercritical region, as judged by comparison with Monte Carlo 
simulation.  At first glance this is surprising, since in the
subcritical regime the stationary state is inactive and might be
regarded as trivial.  Relaxation to this inactive state (and to 
the active state at or near the critical point $p_c$) is however
nontrivial, characterized by stretched exponential,
power-law, or other slowly-converging forms \cite{mannrel}.
It appears to be very difficult to capture such behavior in the kind
of temporal series developed here, which employs a Poissonian
initial distribution.  The reason is that for smaller values of
the particle density ($p < 2$, say), the one-site stationary occupation
distribution $P(n)$ is far from Poissonian.
As $p$ increases, the second factorial moment $\langle n(n-1) \rangle = \rho $
approaches $p^2$, as expected for a Poisson distribution.  Figure 7 shows that
the stationary one-site distribution observed in simulations approaches
the corresponding Poisson distribution with the same density $p$.
(Even for $p=8$ there are significant differences between the distributions; but
analysis of the third and fourth factorial moments suggests
convergence to a Poisson distribution as $p \to \infty$.)

An important open question is whether simply increasing the number of terms would
permit one to analyze the small-$p$ regime.  The present results suggest that
even with 20 or 30 terms this region would remain inaccessible.  
It appears to
be more promising to approach the critical region from above, since for
larger particle densities we find good agreement with numerical results.
In this context it is interesting that the quality of 
predictions near the critical point improves greatly on going from
12 to 16 terms.  This suggests that further extension of the series,
to 20 or more terms, would yield quantitative results for critical
properties, through study of relaxational properties at $p_c$, or
or of stationary properties (as in figure 5) as $p_c$ is
approached from above.  
It would also be of great interest to develop an 
expansion for a stationary property such as $\overline{\rho}_\infty$
directly in powers of the particle density $p$, but this appears to 
be much more difficult than deriving an expansion in powers 
of the time.
We leave such investigations, using modified or extended series, 
as subjects for future work.   
\vspace{1cm}

\noindent{\bf Acknowledgments}

We thank the Referee for a suggestion that allowed us to derive
an extended series.
This work was supported by CNPq, CAPES, and FAPERJ, Brazil.

\newpage 
\begin{table} 
\begin{center} 
\begin{tabular}{|r|r|r|} 
$n$        &  $m$  & $b_{n,m}$ \\         
\hline 
0           &   0      &          1         \\  
\hline
1           &   0      &          1         \\  
\hline   
2           &   0      &          1         \\  
	     &   1      &          8         \\  
\hline   
3           &   0      &          1         \\  
	     &   1      &        66         \\  
	     &   2      &        80         \\  
\hline   
4           &   0      &          1         \\  
	     &   1      &       442         \\  
	     &   2      &  $2 \; 076$         \\  
	     &   3      &       896         \\  
\hline 
5           &   0      &          1         \\  
	     &   1      &  $2 \; 842$         \\  
	     &   2      &  $35 \; 396$         \\  
	     &   3      &  $52 \; 240$              \\  
	     &   4      &  $10 \; 752$              \\  
\hline 
6           &   0      &          1         \\  
	     &   1      &  $18 \; 118$         \\  
	     &   2      &  $516 \; 880$         \\  
	     &   3      & $1 \; 737 \; 952$              \\  
	     &   4      & $1 \; 187 \; 968$               \\  
	     &   5      &  $135 \; 168$              \\  
\hline 
7           &   0      &          1         \\  
	     &   1      &  $\frac{\small 922 \; 265}{\small 8}$         \\  
	     &   2      & $7 \; 040 \; 282$          \\  
	     &   3      & $45 \; 847 \; 512$              \\  
	     &   4      & $67 \; 368 \; 480$               \\  
	     &   5      & $25 \; 614 \; 368$              \\  
	     &   6      & $1 \; 757 \; 184$              \\  
\hline 
8           &   0      &          1         \\  
	     &   1      &  $\frac{\small 5 \; 865 \; 473}{\small 8}$         \\  
	     &   2      &  $\frac{\small 370 \; 752 \; 137}{\small 4}$         \\  
	     &   3      & $1 \; 078 \; 168 \; 434$              \\  
	     &   4      & $2 \; 871 \; 388 \; 040$               \\  
	     &   5      & $2 \; 283 \; 464 \; 832$              \\  
	     &   6      & $536 \; 472 \; 640$              \\  
	     &   7      & $23 \; 429 \; 120$              \\  
\hline 
9           &   0      &          1         \\  
	     &   1      &  $\frac{\small 74 \; 596 \; 747}{\small 16}$         \\  
	     &   2      &  $\frac{\small 4 \; 797 \; 745 \; 191}{\small 4}$         \\  
	     &   3      & $23 \; 841 \; 662 \; 132$              \\  
	     &   4      & $105 \; 679 \; 404 \; 154$               \\  
	     &   5      & $147 \; 137 \; 780 \; 760$              \\  
	     &   6      & $71 \; 353 \; 965 \; 088$              \\  
	     &   7      & $11 \; 072 \; 770 \; 560$              \\  
	     &   8      & $318 \; 636 \; 032$              \\ 
\hline
10          &    0      &         1          \\
	    &    1      &  $\frac{\small 474 \; 336 \; 627}{\small 16}$     \\  
	    &    2      &  $\frac{\small 123 \; 077 \; 063 \; 429}{\small 8}$ \\
	    &    3      &  $\frac{\small 1 \; 018 \; 938 \; 641 \; 745}{\small  2}$  \\
	    &    4      &  $3 \; 584 \; 915 \; 570 \; 625$  \\
	    &    5      &  $7 \; 999 \; 349 \; 570 \; 432$  \\
	    &    6      &  $6 \; 656 \; 488 \; 808 \; 368$  \\
	    &    7      &  $2 \; 121 \; 777 \; 710 \; 528$  \\
	    &    8      &  $227 \; 436 \; 059 \; 136$  \\
	    &    9      &  $4 \; 402 \; 970 \; 624$  \\
\hline
11          &    0      &        1           \\
	    &    1      &  $\frac{\small 12 \; 064 \; 410 \; 263}{\small 64}$  \\
	    &    2      &  $\frac{\small 3 \; 142 \; 928 \; 518 \; 289}{\small 16}$  \\
	    &    3      &  $\frac{\small 85 \; 419 \; 503 \; 179 \; 415}{\small 8}$ \\
	    &    4      &  $116 \; 020 \; 128 \; 091 \; 449$  \\
	    &    5      &  $394 \; 806 \; 480 \; 115 \; 048$  \\
	    &    6      &  $514 \; 548 \; 057 \; 479 \; 072$  \\
	    &    7      &  $278 \; 154 \; 455 \; 793 \; 952$  \\
	    &    8      &  $61 \;313 \; 513 \; 593 \; 600$  \\
	    &    9      &  $4 \; 683 \; 285 \; 856 \; 256$  \\
	    &   10      &  $61 \; 641 \; 588 \; 736$  \\
 \hline            
12          &  0 &       1            \\
	    &  1 &  $\frac{\small 76 \; 711 \; 895 \; 439}{\small 64}$  \\
	    &  2 &  $\frac{\small 80 \; 070 \; 040 \; 225 \; 479}{\small 32}$  \\
	    &  3 &  $\frac{\small 1 \; 770 \; 456 \; 755 \; 814 \; 995}{\small 8}$  \\
	    &  4 &  $\frac{\small 14 \; 610 \; 068 \; 149 \; 248 \; 089}{\small 4}$  \\
	    &  5 &  $18 \; 396 \; 700 \; 126 \; 638 \; 476$  \\
	    &  6 &  $35 \; 650 \; 110 \; 284 \; 461 \; 928$  \\
	    &  7 &  $29 \; 745 \; 976 \; 515 \; 005 \; 712$  \\
	    &  8 &  $11 \; 054 \; 665 \; 928 \; 232 \; 448$  \\
	    &  9 &  $1 \; 747 \; 506 \; 609 \; 502 \; 464$  \\
	    & 10 &  $97 \; 252 \; 577 \; 107 \; 968$  \\
	    & 11 &  $872 \; 465 \; 563 \; 648$  \\
\hline
13          &  0 &      1             \\
	    &  1 &  $\frac{\small 1 \; 951 \; 093 \; 993 \; 893}{\small 256}$
	    \\
	    &  2 &  $\frac{\small 2 \; 037 \; 418 \; 656 \; 354 \; 491}{\small
	    64}$  \\ 
	    &  3 &  $\frac{\small 72 \; 926 \; 486 \; 692 \; 093 \;
	    419}{\small 16}$  \\
	    &  4 &  $\frac{\small 905 \; 058 \; 014 \; 398 \; 112 \;
	    835}{\small 8}$  \\
	    &  5 &  $\frac{\small 1 \; 655 \; 391 \; 460 \; 208 \; 555 \;
	    433}{\small 2}$  \\
	    &  6 &  $2 \; 309 \; 179 \; 626 \; 832 \; 648 \; 726$  \\
	    &  7 &  $2 \; 816 \; 714 \; 002 \; 502 \; 804 \; 952$  \\
	    &  8 &  $1 \; 601 \; 275 \; 099 \; 838 \; 022 \; 656$  \\
	    &  9 &  $426 \; 223 \; 203 \; 786 \; 122 \; 496$  \\
	    & 10 &  $49 \; 655 \; 626 \; 778 \; 919 \; 936$  \\
	    & 11 &  $2 \; 046 \; 635 \; 410 \; 882 \; 560$  \\
	    & 12 &  $12 \; 463 \; 793 \; 766 \; 400$  \\  
\hline
14          &  0 &      1             \\
	    &  1 &  $\frac{\small 2 \; 613 \; 736 \; 799 \; 297}{\small 64}$
	    \\
	    &  2 &  $\frac{\small 11 \; 934 \; 019 \; 637 \; 184 \;
	    639}{\small 32}$  \\ 
	    &  3 &  $\frac{\small 711 \; 799 \; 150 \; 376 \; 749 \;
	    517}{\small 8}$  \\
	    &  4 &  $\frac{\small 53 \; 725 \; 847 \; 102 \; 644 \; 113 \;
	    051}{\small 16}$  \\ 
	    &  5 &  $\frac{\small 142 \; 451 \; 880 \; 202 \; 934 \; 178 \;
	    839}{\small 4}$  \\
	    &  6 &  $140 \; 971 \; 191 \; 603 \; 315 \; 396 \; 510$  \\
	    &  7 &  $243 \; 993 \; 717 \; 303 \; 561 \; 711 \; 492$  \\
	    &  8 &  $201 \; 186 \; 302 \; 850 \; 192 \; 322 \; 944$  \\
	    &  9 &  $81 \; 746 \; 928 \; 038 \; 823 \; 569 \; 408$  \\
	    & 10 &  $16 \; 113 \; 035 \; 142 \; 846 \; 829 \; 824$  \\
	    & 11 &  $ 1 \; 415 \; 730 \; 263 \; 534 \; 155 \; 776$  \\
	    & 12 &  $ 43 \; 811 \; 063 \; 460 \; 921 \; 344$  \\
	    & 13 &  $179 \; 478 \; 630 \; 236 \; 160$  \\    
\hline
15          &  0 &      1             \\
	    &  1 &  $\frac{\small 531 \; 822 \; 407 \; 449 \; 409}{\small
	    2048}$  \\
	    &  2 &  $\frac{\small 606 \; 748 \; 047 \; 325 \; 325 \;
	    193}{\small 128}$  \\
	    &  3 &  $\frac{\small 116 \; 603 \; 968 \; 592 \; 784 \; 196 \;
	    927}{\small 64}$  \\
	    &  4 &  $\frac{\small 819 \; 559 \; 563 \; 865 \; 455 \; 675 \;
	    379}{\small 8}$  \\
	    &  5 &  $\frac{\small 12 \; 371 \; 190 \; 775 \; 573 \; 187 \; 034
	    \; 899}{\small 8}$  \\
	    &  6 &  $8 \; 499 \; 377 \; 166 \; 784 \; 889 \; 887 \; 638$  \\
	    &  7 &  $20 \; 286 \; 352 \; 375 \; 993 \; 324 \; 998 \; 496$  \\
	    &  8 &  $23 \; 298 \; 464 \; 567 \; 721 \; 533 \; 566 \; 328$  \\
	    &  9 &  $13 \; 579 \; 307 \; 980 \; 015 \; 469 \; 945 \; 184$  \\
	    & 10 &  $4 \; 068 \; 154 \; 005 \; 082 \; 098 \; 401 \; 408$  \\
	    & 11 &  $607 \; 110 \; 051 \; 667 \; 479 \; 652 \; 352$  \\
	    & 12 &  $40 \; 856 \; 912 \; 571 \; 394 \; 580 \; 480$  \\
	    & 13 &  $957 \; 525 \; 442 \; 027 \; 462 \; 656$  \\
	    & 14 &  $2 \; 602 \; 440 \; 138 \; 424 \; 320$  \\        
\hline
16          &  0 &      1             \\
	    &  1 &  $\frac{\small 422 \; 699 \; 161 \; 810 \; 361}{\small
	    256}$  \\
	    &  2 &  $\frac{\small 61 \; 687 \; 281 \; 835 \; 997 \; 869 \;
	    817}{\small 1024}$  \\
	    &  3 &  $\frac{\small  1 \; 192 \; 545 \; 870 \; 991 \; 479 \; 699
	    \; 681}{\small 32}$  \\
	    &  4 &  $\frac{\small 12 \; 465 \; 559 \; 773 \; 385 \; 531 \; 628
	    \; 949}{\small 4}$  \\
	    &  5 &  $\frac{\small 133 \; 184 \; 916 \; 649 \; 036 \; 025 \;
	    384 \; 179}{\small 2}$  \\
	    &  6 &  $502 \; 866 \; 857 \; 598 \; 243 \; 404 \; 511 \; 546$  \\
	    &  7 &  $ 1 \; 627 \; 359 \; 034 \; 988 \; 855 \; 536 \; 002 \;
	    199$  \\
	    &  8 &  $2 \; 537 \; 319 \; 446 \; 210 \; 036 \; 202 \; 445 \;
	    148$  \\
	    &  9 &  $2 \; 040 \; 132 \; 355 \; 769 \; 944 \; 077 \; 918 \;
	    400$  \\
	    & 10 &  $872 \; 811 \; 268 \; 389 \; 569 \; 306 \; 302 \; 976$  \\
	    & 11 &  $198 \; 169 \; 235 \; 678 \; 101 \; 485 \; 620 \; 992$  \\
	    & 12 &  $22 \; 853 \; 161 \; 358 \; 227 \; 259 \; 040 \; 768$  \\
	    & 13 &  $1 \; 195 \; 547 \; 596 \; 367 \; 062 \; 589 \; 440$  \\
	    & 14 &  $21 \; 401 \; 594 \; 847 \; 260 \; 721 \; 152$  \\
	    & 15 &  $37 \; 965 \; 009 \; 078 \; 190 \; 080$  \\
\end{tabular} 
\end{center} 
\label{tab1} 
\noindent{\sf Table I. Series coefficients in the expansion of the activity.}
\end{table}

\newpage
\noindent FIGURE CAPTIONS 
\vspace{1cm} 
 
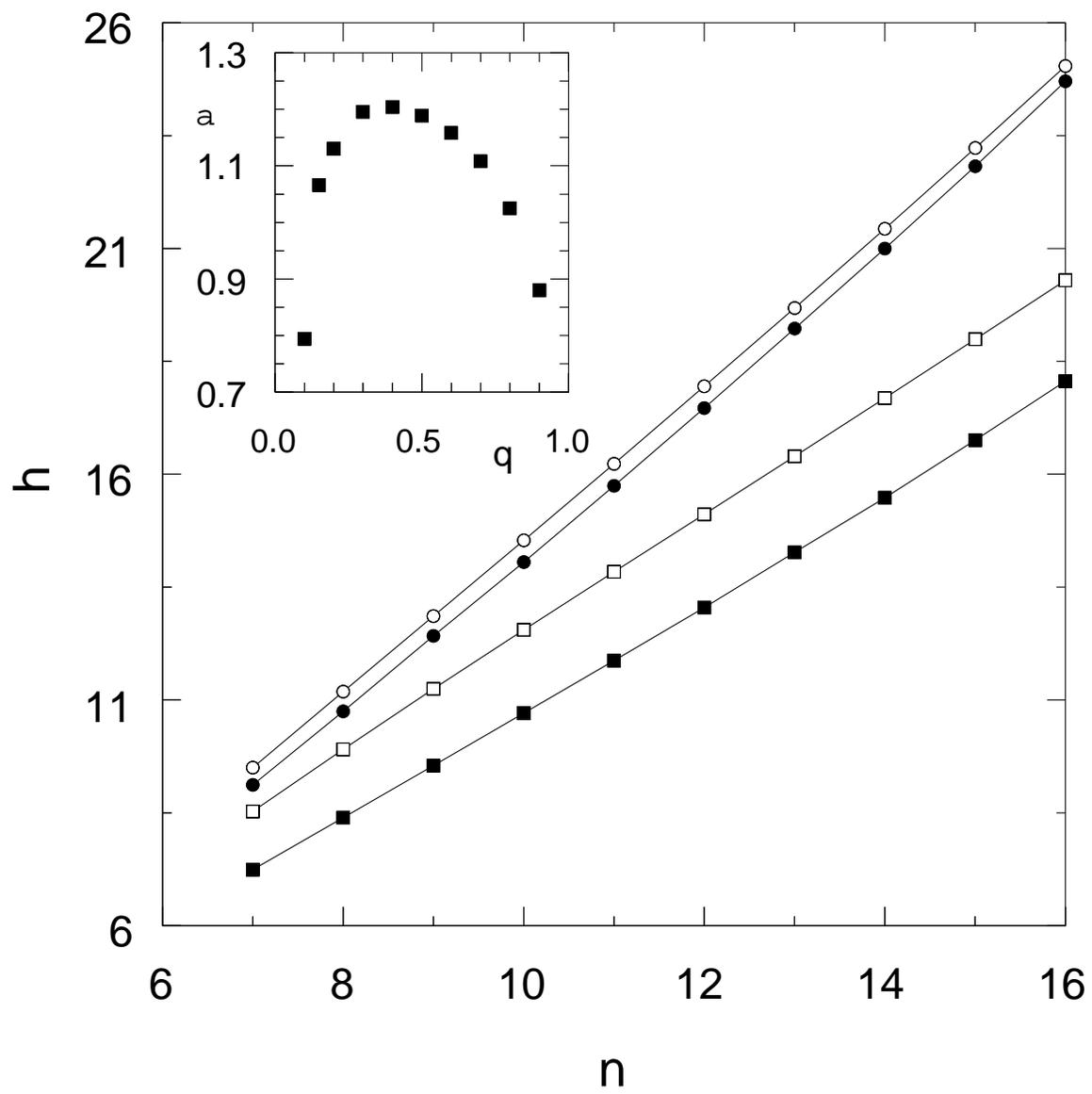
\noindent Figure 1. Function $h_n(q)$ as defined in text, for 
$q=0.2$ ($\blacksquare$);
$q=0.4$ ($\bullet$);
$q=0.6$ ($\circ$);
$q=0.8$ ($\Box$).  Observe that $h$ grows faster than linearly
for $q=0.2$, 0.4.  Inset: growth exponent $\alpha (q)$ defined
via $h_n(q) \sim n^{\alpha(q)}$.
\vspace{1em}

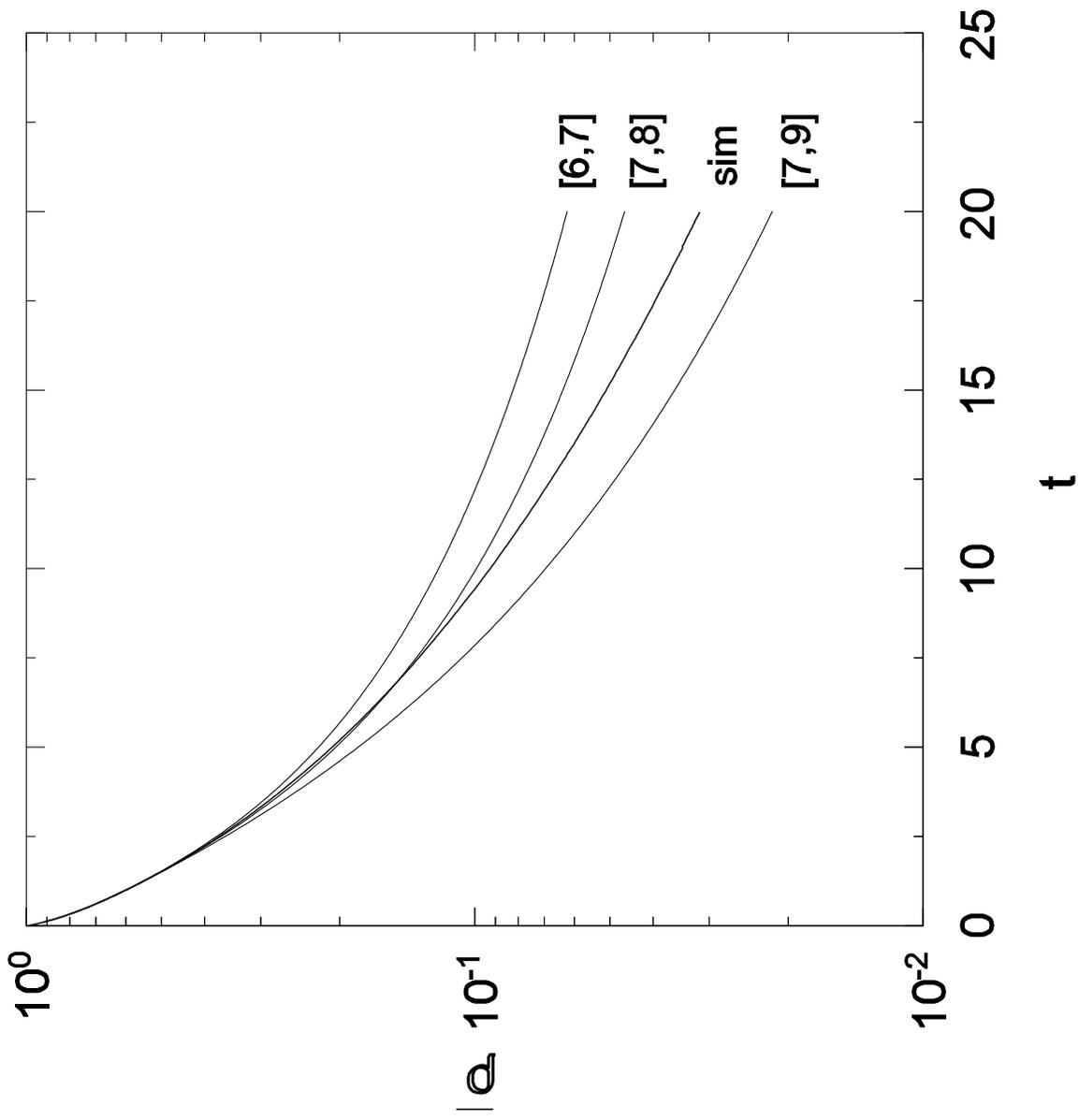
\noindent Figure 2.  Normalized activity 
$\overline{\rho} \equiv \rho(t)/p^2$ versus time for $p=1/2$ 
from simulation and various  
Pad\'e approximants to the times series as indicated.
\vspace{1em} 

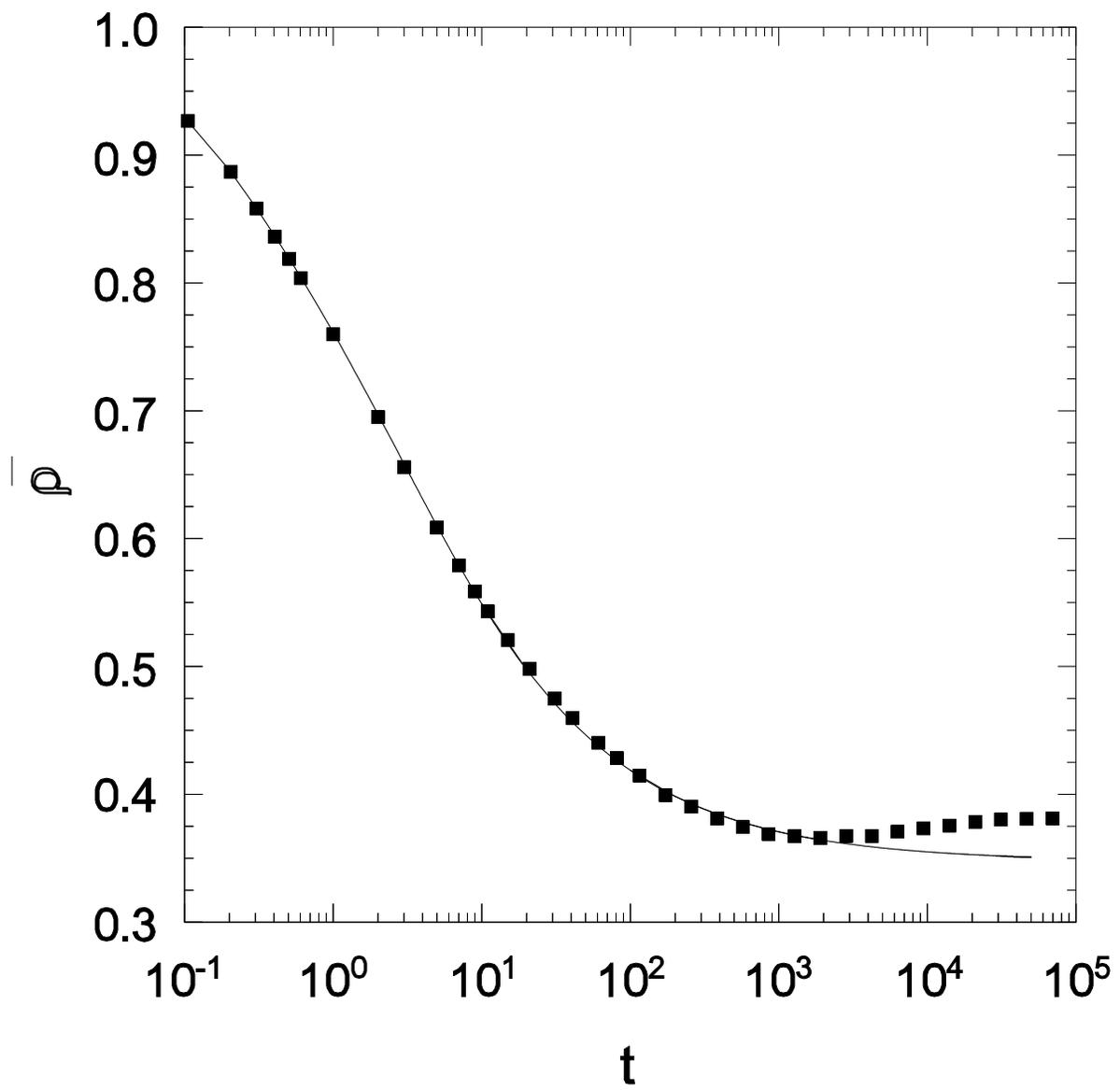
\noindent Figure 3.  Normalized activity for $p=1$. 
Symbols: simulation result; curve: series prediction as described in text.
\vspace{1em}

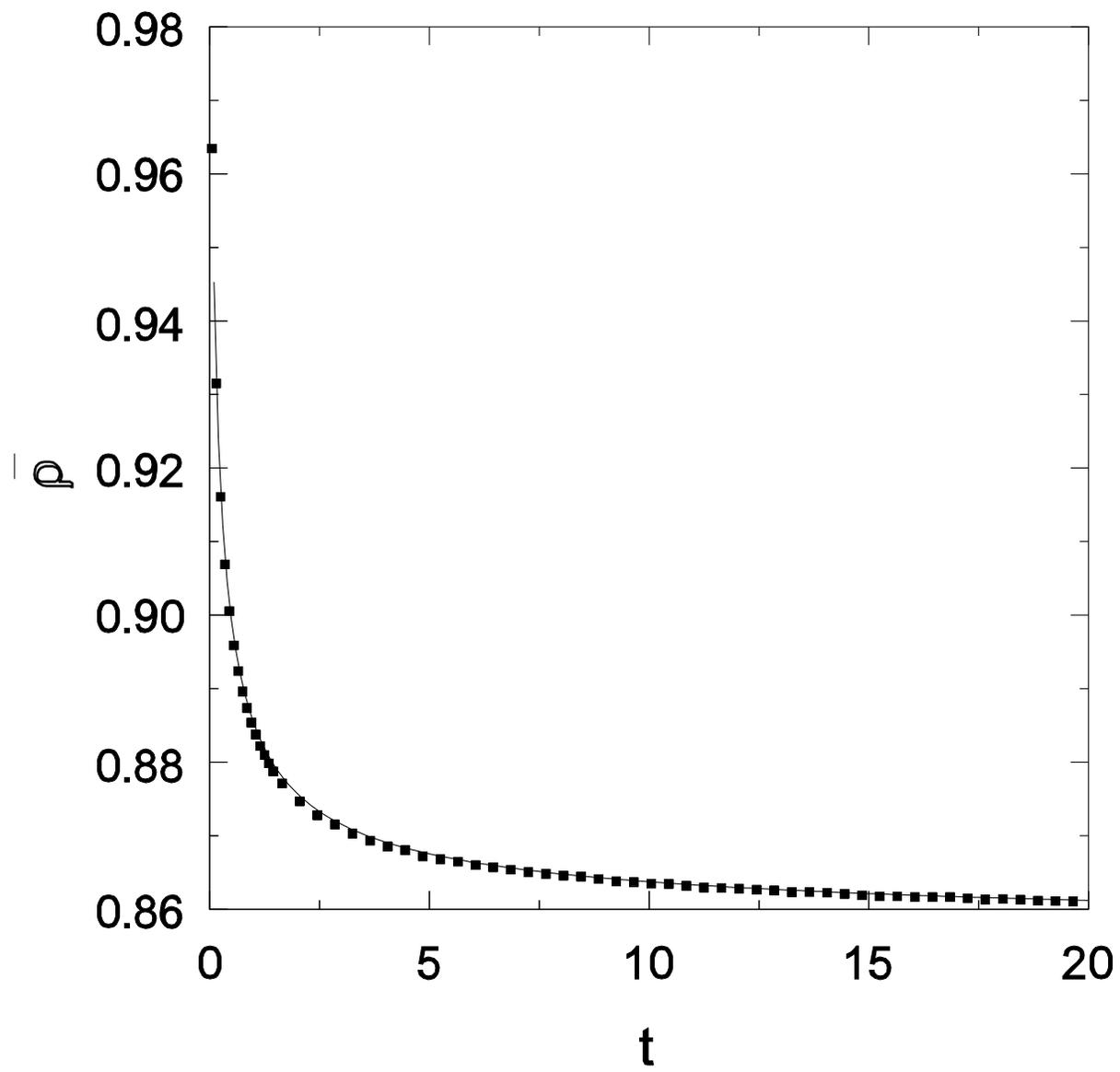
\noindent Figure 4.  As in figure 3 for but for $p=2$.
\vspace{1em} 
 
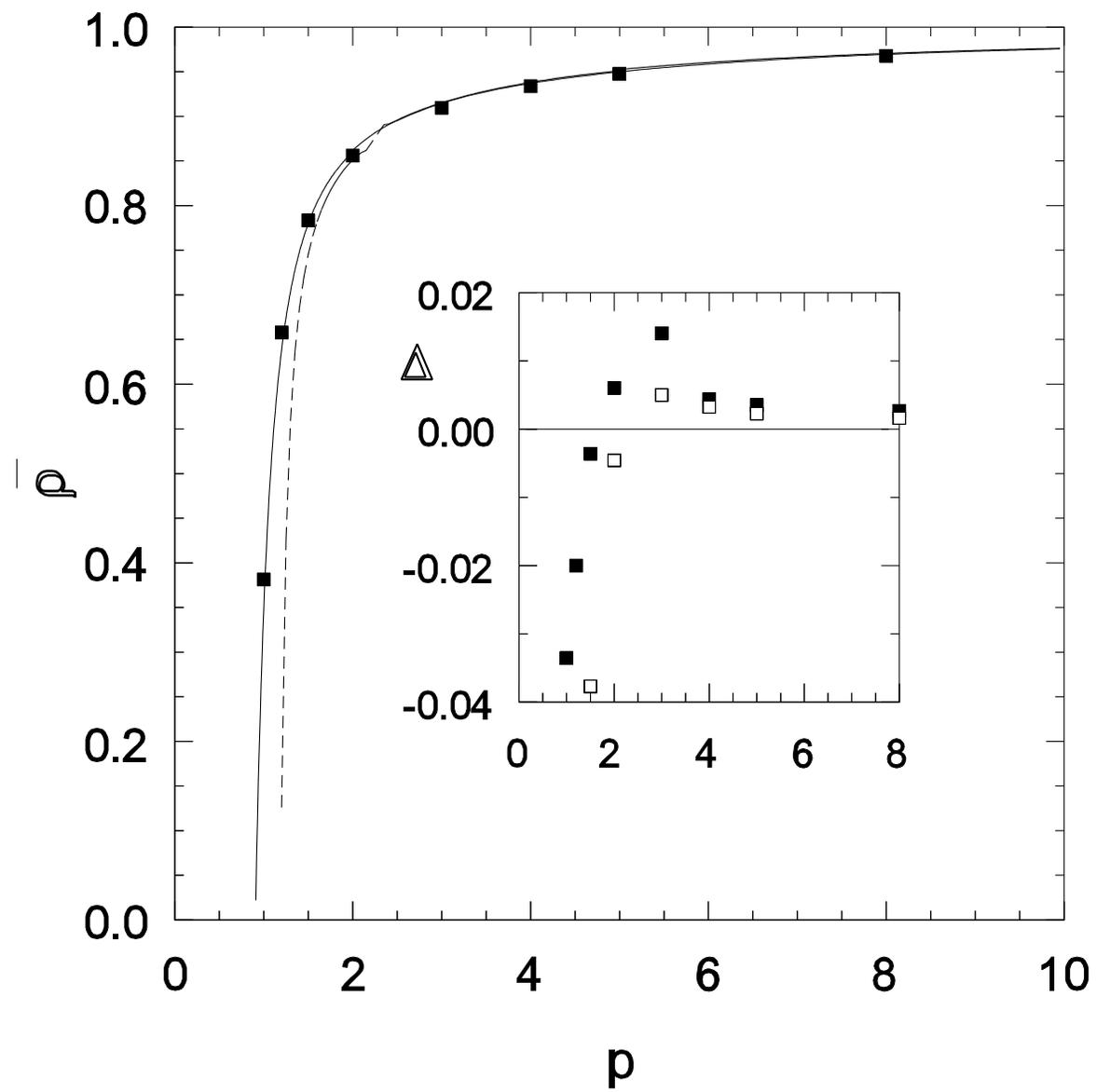
\noindent Figure 5. Main graph: limiting activity $\overline{\rho}_\infty$
versus particle density $p$.  Points: simulation; solid
curve: series prediction using transformation (\ref{gamma}) with $b=0.57$,
[8,8] Pad\'e approximant; dashed line: same approximant and transformation
but using $b=5$.
Inset: difference $\Delta = \overline{\rho}_{\infty,series} -
\overline{\rho}_{\infty,sim}$ for $b=0.57$ ($\blacksquare$) and
$b=5$ ($\Box$).
\vspace{1em} 

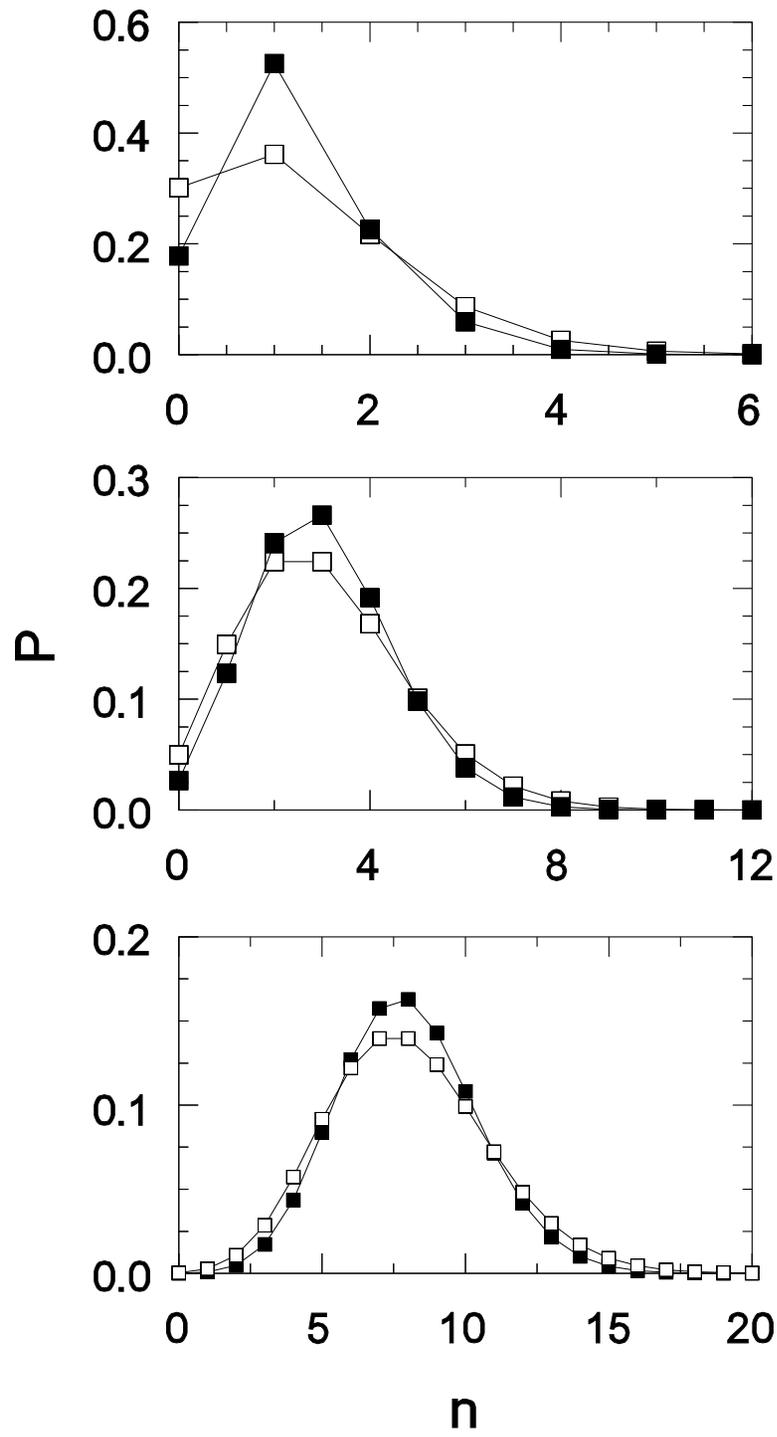
\noindent Figure 6.  Single-site occupancy distributions $P(n)$
obtained in simulation ($\blacksquare$) compared with the
corresponding Poisson distribution ($\Box$).  Upper panel:
$p=1.2$; middle: $p=3$; lower: $p=8$.

\end{document}